\documentclass[%
aps,
prb,
reprint,
superscriptaddress,
showkeys,
preprintnumbers,
amsmath,amssymb,
floatfix,
]{revtex4-1}

\usepackage{graphicx}
\usepackage{siunitx}
\usepackage{dcolumn}
\usepackage{bm}

\begin{document}
	
	\preprint{Version 9}
	
	\title{Magnons in a Quasicrystal: Propagation, Extinction and Localization of Spin Waves in Fibonacci Structures}
	
	\author{Filip Lisiecki}
	\email{flisiecki@ifmpan.poznan.pl}
	\thanks{F.L. and J.R. contributed equally to this work.}
	\affiliation{Institute of Molecular Physics, Polish Academy of Sciences, Pozna\'n, Poland}
	
	\author{Justyna Rych\l{}y}
	\email{rychly@amu.edu.pl}
	\affiliation{Faculty of Physics, Adam Mickiewicz University, Pozna\'n, Poland}
	
	\author{Piotr Ku\'swik}
	\affiliation{Institute of Molecular Physics, Polish Academy of Sciences, Pozna\'n, Poland}
	\affiliation{Centre for Advanced Technologies, Adam Mickiewicz University, Pozna\'n, Poland}
	
	\author{Hubert G\l{}owi\'nski}
	\affiliation{Institute of Molecular Physics, Polish Academy of Sciences, Pozna\'n, Poland}
	
	\author{Jaros\l{}aw W. K\l{}os}
	\affiliation{Faculty of Physics, Adam Mickiewicz University, Pozna\'n, Poland}
	\affiliation{Institute of Physics,University of Greifswald, Greifswald, Germany}
	
	\author{Felix~Gro\ss}
	\affiliation{Max Planck Institute for Intelligent Systems, Stuttgart, Germany}
	
	\author{Nick Tr\"ager}
	\affiliation{Max Planck Institute for Intelligent Systems, Stuttgart, Germany}
	
	\author{Iuliia Bykova}
	\affiliation{Max Planck Institute for Intelligent Systems, Stuttgart, Germany}
	
	\author{Markus Weigand}
	\affiliation{Max Planck Institute for Intelligent Systems, Stuttgart, Germany}
	
	\author{Mateusz Zelent}
	\affiliation{Faculty of Physics, Adam Mickiewicz University, Pozna\'n, Poland}
	
	\author{Eberhard J. Goering}
	\affiliation{Max Planck Institute for Intelligent Systems, Stuttgart, Germany}
	
	\author{Gislea Sch\"utz}
	\affiliation{Max Planck Institute for Intelligent Systems, Stuttgart, Germany}
	
	\author{Maciej Krawczyk}
	\affiliation{Faculty of Physics, Adam Mickiewicz University, Pozna\'n, Poland}
	
	\author{Feliks Stobiecki}
	\affiliation{Institute of Molecular Physics, Polish Academy of Sciences, Pozna\'n, Poland}
	
	\author{Janusz Dubowik}
	\affiliation{Institute of Molecular Physics, Polish Academy of Sciences, Pozna\'n, Poland}
	
	\author{Joachim Gr\"afe}
	\email{graefe@is.mpg.de}
	\affiliation{Max Planck Institute for Intelligent Systems, Stuttgart, Germany}
	
	\date{\today}
	
	\begin{abstract}
		Magnonic quasicrystals exceed the possibilities of spin wave (SW) manipulation offered by regular magnonic crystals, because of their more complex SW spectra with fractal characteristics. Here, we report the direct X-ray microscopic observation of propagating SWs in a magnonic quasicrystal, consisting of dipolar coupled permalloy nanowires arranged in a one-dimensional Fibonacci sequence. SWs from the first and second band as well as evanescent waves from the band gap between them are imaged. Moreover, additional mini-band gaps in the spectrum are demonstrated, directly indicating an influence of the quasiperiodicity of the system. Finally, the localization of SW modes within the Fibonacci crystal is shown. The experimental results are interpreted using numerical calculations and we deduce a simple model to estimate the frequency position of the magnonic gaps in quasiperiodic structures. The demonstrated features of SW spectra in one-dimensional magnonic quasicrystals allows utilizing this class of metamaterials for magnonics and makes them an ideal basis for future applications.
	\end{abstract}
	
	
	\maketitle
	
	
	\section{Introduction}
	Magnonic crystals are periodically modulated magnetic structures, which enable tailoring of the magnonic band structure and formation of allowed and forbidden bands in the spin wave (SW) spectrum~\cite{Krawczyk2014Review,Stamps20142014}. Additionally, the SW spectrum can easily be modified by an external magnetic field or a change of the magnetization configuration in the structure~\cite{Tacchi2010Analysis}. This offers fine tuning and re-programmability~\cite{Topp2010Making}, which are desirable properties for potential applications~\cite{Chumak2015Magnon}. Apart from periodic modulations, defects in the regular structure can also alter SW propagation. Defects can cause the appearance of localized SW modes with their amplitude confined to the disturbed area. Equally, they can introduce additional magnonic branches at frequencies inside the band gaps of the SW spectrum~\cite{Di2014Band,Rychly2017Spin,Kruglyak2006Spin,Klos2013Symmetry}. In a periodically arranged array of magnetic stripes, a stripe with differing dimensions can be considered and designed as such a defect. Thereby, providing an opportunity for the design of cavity resonators and ultra-narrow band filters~\cite{Zhang2016Spin-wave,Filimonov2012Magnetostatic}.
	
	Quasicrystals, unlike crystals, do not have periodicity albeit possessing long-range order, which results in a discrete diffraction pattern~\cite{Jagodzinski1991Diffraction}. They are characterized by more complex dispersion relations than those of periodic systems with an increased number of forbidden band gaps and narrow allowed bands. Spectra of quasicrystals also include localized excitations concentrated in various parts of the self-similar structure, which is a characteristic feature of such aperiodic systems~\cite{2009Metamaterials}. Previously, these properties have been widely studied in the context of photonics and phononics for one-dimensional (1D) quasicrystals designed by using the Fibonacci sequence~\cite{Macia1998Optical,Golmohammadi2007Narrowband,Lubin2013Quasiperiodic}.
	
	Quasiperiodicity has also been applied to magnonic systems~\cite{Costa2011Partial,Coelho2011Transmission,Costa2013Band,Coelho2010Quasiperiodic,Machado2013Static,Rychly2015Spin,Rychly2016Spin,Grishin2013Dissipative,Bhat2013Controlled,Bhat2014Ferromagnetic,Farmer2015Magnetic,Choudhury2017Efficient,Topp2009Interaction}. Fundamental theoretical studies used Fibonacci structures comprised of multilayers~\cite{Costa2011Partial,Coelho2011Transmission,Costa2013Band,Coelho2010Quasiperiodic,Machado2013Static} or bi-component stripes~\cite{Rychly2015Spin,Rychly2016Spin} to investigate the spectrum of exchange or dipolar SWs respectively. However, these hypothetical structures were far from experimental realization. First experimental realizations of quasiperiodicity in magnonics were based on groves in a micrometre thick YIG film~\cite{Grishin2013Dissipative}. All electrical spin wave spectroscopy measurements indicated an influence of the quasiperiodicity on the SW spectrum~\cite{Grishin2013Dissipative}. A key experimental realization of flexible magnonic quasicrystals were 2D arrays of Py nanobars in Penrose, Ammann, or Fibonacci arrangement~\cite{Bhat2013Controlled,Bhat2014Ferromagnetic,Farmer2015Magnetic,Choudhury2017Efficient,Topp2009Interaction}. However, these studies were focused on the magnetization reversal processes and only demonstrated a rich ferromagnetic resonance (FMR) spectrum of these systems, hinting at a more complex SW band structure~\cite{Grishin2013Dissipative,Bhat2013Controlled,Bhat2014Ferromagnetic,Farmer2015Magnetic,Choudhury2017Efficient,Topp2009Interaction}. So far, there has been no study of the microscopic behaviour, propagation, nor localization of SWs in magnonic quasicrystals.
	
	\begin{figure}
		\includegraphics[width=0.66\columnwidth]{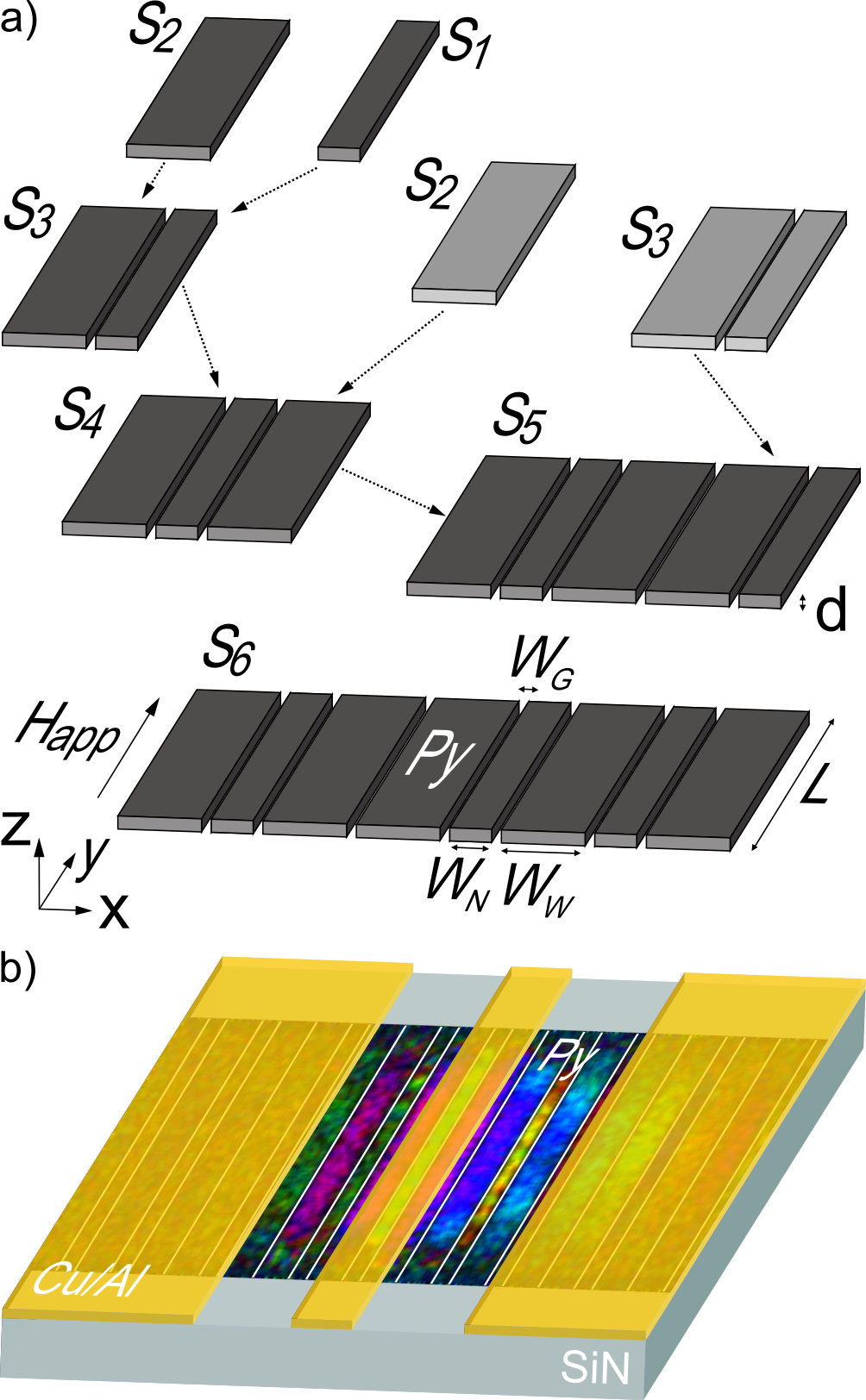}
		\caption{\emph{(a)} Arrangement of the first sequences ($S_1$ to $S_6$) of Fibonacci structures consisting of Py NWs separated by air gaps. \emph{(b)} Sample scheme with CPW (yellow) used for SW excitation and superimposed STXM measurement.\label{sketch}}
	\end{figure}
	
	Here, we present the direct microscopic observation of propagating SWs in a 1D quasiperiodic magnonic structure formed by thin Py stripes arranged in a Fibonacci sequence. This is achieved by using scanning transmission X-ray microscopy (STXM) that provides ultimate spatial ($<20$ nm) and temporal ($<50$ ps) resolution. Propagating modes from the first and second bands were detected with a band gap between them. Furthermore, the existence of a mini-band gap within the first band was demonstrated, showing the influence of the quasiperiodicity on the dispersion relation. We complement the experimental results with calculations that show good agreement with the resonant frequencies and profiles of the investigated modes and allow their interpretation. Moreover, we propose a simple model based on diffraction structure factor calculations, which allows predicting the frequency positions of most magnonic bandgaps in the SW spectra of 1D magnonic quasicrystals.
	
	\section{Methods}
	
	Therefore, narrow ($W_N = 700$ nm) and wide ($W_W = 2W_N = 1400$ nm) NWs were arranged in arrays with quasiperiodic order according to Fibonacci's inflation rule. The total width of these arrays was 100 $\mu$m. According to Fibonacci's inflation rule, a sequence of higher order $n$ is determined by the concatenation of the two previous structures ($S_n = S_{n-1} + S_{n-2}$) as shown in Figure~\ref{sketch}a. To ensure magnetostatic coupling between the NWs an air gap ($W_G$ = 80 nm) was introduced between adjacent NWs~\cite{Topp2009Interaction}. A static magnetic field $H_\text{app}$ was applied parallel to the NWs axis (y-axis). A sketch of the sample geometry superimposed with an exemple of a STXM result, indicating SW propagation across the NW array, is shown in Figure~\ref{sketch}b. 
	
	\subsection{Sample Preparation}
	Nanowires (NWs, length $L = 10$ $\mu$m) were fabricated by e-beam lithography and subsequent lift-off in a thin Py film (Ni\textsubscript{80}Fe\textsubscript{20}, thickness $d = 30$ nm) on a Si\textsubscript{3}N\textsubscript{4}(100 nm)/Si substrate with membrane windows for X-ray transmission measurements. 
	
	For SW excitation a coplanar waveguide (CPW) made from Cu(150 nm)/Al(10 nm) with a 2 $\mu$m wide signal line was fabricated on top of the structure using direct laser lithography and a lift-off technique~\cite{Koczorowski2017CMOS-}. The CPW lines are aligned along the NWs axis to excite a dynamic magnetic field with a component perpendicular to the array (x-axis). 
	
	\subsection{X-Ray Microscopy}
	Time-resolved STXM measurements were conducted at the MPI IS operated MAXYMUS end station at the UE46-PGM2 beam line at the BESSY II synchrotron radiation facility. The samples were illuminated under perpendicular incidence by circularly polarized light in an applied in-plane field of up to 240 mT that was generated by a set of four rotatable permanent magnets~\cite{Nolle2012Note}. The photon energy was set to the absorption maximum of the Fe $L_3$ edge to get optimal XMCD contrast for imaging. A lock-in like detection scheme allows sample excitation at arbitrary frequencies at a time resolution of 50 ps using all photons emitted by the synchrotron. Phase and amplitude of the SWs is extracted by a temporal Fourier transformation per pixel~\cite{Gross2019Nanoscale}.
	
	\subsection{Numerical Calculations}
	To calculate SW spectra we solve the Landau-Lifshitz (LL) equation:
	\begin{equation}
	\frac{\partial \textbf{M}(\textbf{r},t)}{\partial t}=\mu_0\gamma \textbf{M}(\textbf{r},t)\times \textbf{H}_\text{eff}(\textbf{r},t),\label{eq1}
	\end{equation}
	where $t$ is time and $\textbf{r}$ is the position vector. Damping has been neglected in the calculations. We expressed the effective magnetic field $\textbf{H}_\text{eff}(\textbf{r},t)$ as a sum of three terms: $\textbf{H}_\text{eff}(\textbf{r},t)=\textbf{H}_\text{app}+\textbf{H}_\text{ex}(\textbf{r},t)+\textbf{H}_\text{dm}(\textbf{r},t)$. We assumed that the external field: $\textbf{H}_\text{app}$, applied along the NW, is constant in time and homogeneous in space. In considered model (see Figure~\ref{sketch}), the magnetic configuration is saturated and the static magnetization is oriented along the axes of infinitely long NWs. Therefore, both  the exchange field $\textbf{H}_\text{ex}(\textbf{r},t)$ and dipolar field $\textbf{H}_\text{dm}(\textbf{r},t)$  do not contain static components. The dynamic components of $\textbf{H}_\text{ex}(\textbf{r},t)$ and $\textbf{H}_\text{dm}(\textbf{r},t)$ are defined in Ref.~\cite{Mruczkiewicz2013Standing}.
	
	From Equation~\ref{eq1} we find the dynamical components of the magnetization $\textbf{m}(\textbf{r},t)$ where $\textbf{M}(\textbf{r},t)=M_z(\textbf{r})\hat{e}_z+\textbf{m}(\textbf{r},t)$. We use a linear approximation, \textit{i.e.} we neglect the higher order terms arising in Equation~\ref{eq1} with respect to $m$. This is justified when $M_z$ is assumed to be constant in time, namely when $|m(r,t)|\ll M_z (r)$, and therefore $M_z\approx M_S$, where $M_S$ is saturation magnetization. We seek solutions of the LL Equation~\ref{eq1} in the form of monochromatic SWs, having harmonic dynamics in time: $e^{2\pi ift}$, where $f$ is the frequency of the SW. Equation~\ref{eq1} is complemented with the Maxwell equations to determine the demagnetizing fields. With these equations we define the eigenvalue problem, which is solved by using a FEM approach with COMSOL 5.1 to obtain the dispersion relation and profiles of the SWs. For more details concerning this computation, we refer to Ref.~\cite{Gurevich1996Magnetization}. In numerical calculation we used saturation magnetization $M_S=0.76 \times 10^6$  A/m, the exchange constant $A=1.3 \times 10^{-11}$  J/m, and gyromagnetic ratio $\gamma=1.76\times 10^{11}$ rad/sT for Py.
	
	\section{Results \& Discussion}
	
	\begin{figure*}[t]
		\includegraphics[width=\textwidth]{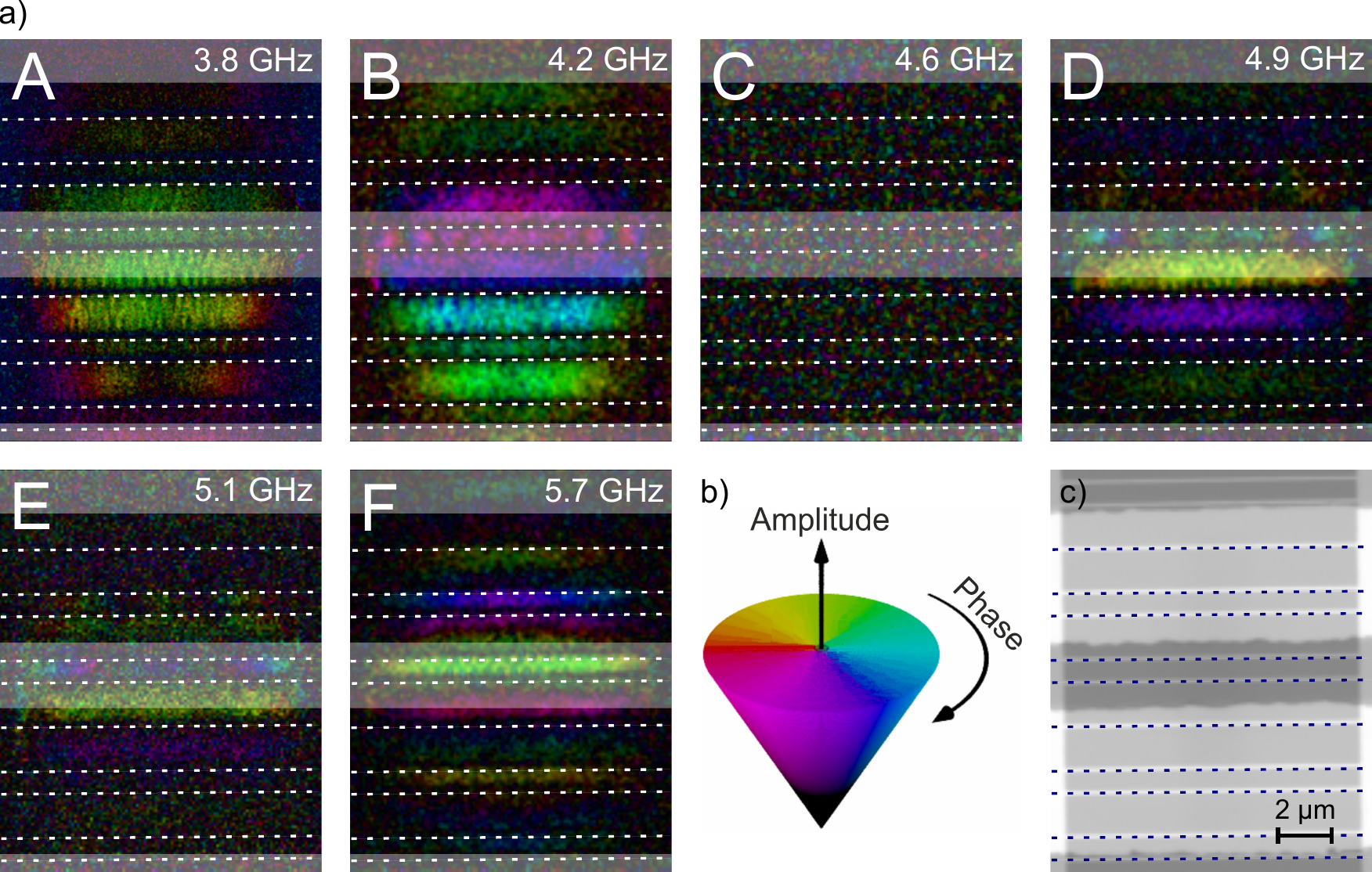}
		\caption{\emph{(a)} SW amplitude and phase for different excitation frequencies at 5 mT. The transparent gray rectangles mark the position of the CPW and the dashed white lines mark the gaps between the stripes. \emph{(b)} Color code for SW amplitude (brightness) and phase (color). \emph{(c)} Static image of the Fibonacci structure (light gray) with the signal line (dark gray) near the center of the image, and the ground lines at the top and bottom edges of the image (dark gray). See Supporting Information for the corresponding videos.\label{stxm}}
	\end{figure*}

	\begin{figure*}[t]
		\includegraphics[width=\textwidth]{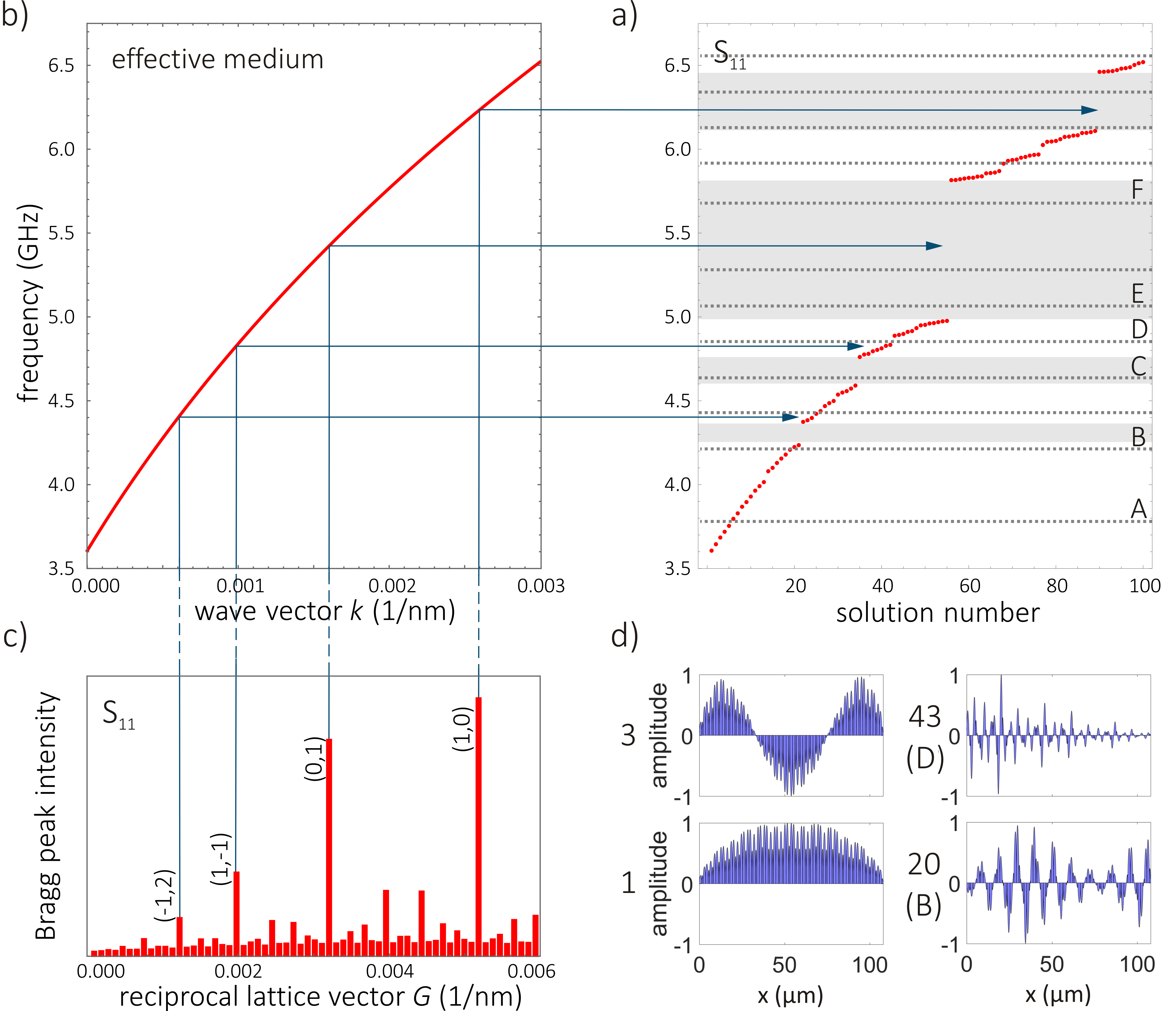}
		\caption{Results of the numerical calculations: \emph{(a)} the SW spectrum for the planar magnonic quasicrystal is compared to \emph{(b)} the analytical SW dispersion in a homogeneous plain film with effective material parameters. In subfigure \emph{(a)} the frequencies of the eigenmodes of the $S_{11}$ Fibonacci sequence (\textit{cf.} Figure 1) are sorted in ascending order (solid red dots). The magnonic gaps are marked as gray areas and the horizontal dashed lines indicate the frequencies at which the experimental measurements have been performed. The horizontal lines marked from A to F in \emph{(a)} indicate the modes presented in Figure~\ref{stxm}a. The spatial Fourier spectrum of the spatially dependent material parameter (\textit{e.g}. $M_S$) was calculated for the considered $S_{11}$ sequence. The 1D reciprocal lattice vectors $G$ of the Fourier components determine the SW’s wave numbers $k$ for which the Bragg condition ($G/2 = k$) is fulfilled. The blue arrows link the: \emph{(c)} reciprocal lattice vectors $G$, corresponding to the Bragg peaks of highest intensity to \emph{(b)} the frequencies of the SWs in the effective medium, which then point at \emph{(a)} the largest magnonic gaps of the Fibonacci quasicrystal. The integer numbers in brackets over the highest Bragg peaks at \emph{(c)} are the pairs of indexes for 1D reciprocal vectors (\textit{cf.} Equation~\ref{eqa2} in the appendix) of the Fibonacci quasicrystal. At \emph{(d)} the amplitude and phase distributions of some modes are shown. The modes with solution number 20 and 43 are compared with the measured modes B and D respectively.\label{calculation}}
	\end{figure*}
	
	Time-resolved STXM measurements with continuous wave excitation from the CPW were conducted in the frequency range from 3.6 to 6.6 GHz (\textit{cf.} Figure~\ref{stxm}). Magnetization dynamics from selected resonances (A-F) in an applied static field of $H_\text{app} = 5$ mT are shown in Figure~\ref{stxm}a. In these images, the SW amplitude is indicated as brightness and the relative SW phase as color (\textit{cf.} Figure~\ref{stxm}b). Additionally, the static Fibonacci structure (light gray) with the CPW (dark gray) is shown in Figure~\ref{stxm}c.
	
	To complement the experimental results numerical calculations were perfromed. In these calculations we assumed an array of infinitely long Py NWs, keeping the thickness, widths and the Fibonacci arrangement (\textit{cf.} Figure~\ref{sketch}a) from the experiment. We considered $S_{11}$ Fibonacci sequence composed of 55 wide and 34 narrow NWs (\textit{cf.} Figure~\ref{sketch}). The calculated eigenfrequencies of the SW modes are shown in Figure~\ref{calculation}a as full red dots. The horizontal axis in Figure~\ref{calculation}a shows the solution number, indicating the assumed order of SW modes. We are showing the first 100 solutions with the lowest frequencies. It is worth to note that the number of modes in the sub-bands shown in Figure~\ref{calculation}a is also equal to one of the Fibonacci numbers: 21 $-$ ($3.6-4.25$ GHz), 13 $-$ ($4.35-4.6$ GHz), 21 $-$ ($4.75-5.0$ GHz) 34 $-$ ($5.8-6.1$ GHz). The magnonic gaps are marked as gray areas and the horizontal dashed lines indicate the frequencies at which the experimental measurements have been performed. The horizontal lines marked from A to F in Figure~\ref{calculation}a indicate the modes presented in Figure~\ref{stxm}a.
	
	\subsection{Propagating Spin Waves}
	
	For all frequencies where SWs could be excited in the experiment, a gradual variation of the phase along the propagation direction is visible in the measurements, represented by a gradual variation of color in Figure~\ref{stxm}a. This indicates that the SWs in the Fibonacci structure exhibit a propagating character. For 3.8 GHz (mode A in Figure~\ref{stxm}a) and 4.2 GHz (mode B), SW modes with strong amplitude are visible. At higher frequencies (modes D$-$F) we again observe SW excitation. For 4.9 GHz (mode D) and 5.1 GHz (mode E) SWs are visible, but at a much weaker amplitude that decays rapidly with distance from the signal line which indicates purely forced oscillation below the stripline. However, for even higher frequencies, \textit{i.e.} 5.7 GHz (mode F), there is again a strong signal for SW excitation. In the higher frequency band (modes D$-$F) the phase varies over a shorter distance, indicating a smaller wavelength of the excited SWs.
	
	In calculations, the spectrum starts at 3.6 GHz, which is significantly above the FMR frequency of 1.9 GHz for Py. This significant upshift can be attributed to the dipolar pinning of SWs~\cite{Guslienko2002Effective} on the interfaces of the ferromagnetic stripes that are separated by air gaps resulting from dynamic demagnetization (\textit{cf.} appendix). In the spectrum (\textit{cf.} Figure~\ref{calculation}a), we can identify a wide band gap ranging from 5.0 to 5.8 GHz (indicated by gray horizontal bar), that splits the spectrum into two parts, \textit{i.e}. the two main bands. Below this gap, there are 55 solutions, which is equal to the number of wider NWs in the whole simulated structure. We observed the same number of solutions in the bands of periodic structure (magnonic crystal) composed of 55 unit cells where each cell contains one wide and one narrow NW of the same dimensions as in considered Fibonacci sequence.
	
	This indicates that the strong excitations (modes A, B, D and F) lie in allowed bands (first and second band respectively), while the forced oscillation (mode E) lies in a band gap. Overall, the results of the calculations match the measured data very well, validating our interpretation.
	
	From the measurements, we are able to estimate the decay length ($L$) of SWs propagating through the structure. This is determined from an exponential fit of the spatial amplitude distribution according to $Ae^{-x/\Lambda}$, where $A$ is the SW amplitude and $x$ is the distance from the excitation source, \textit{i.e.} the stripline. We calculated the decay length to be $\Lambda = 14\pm2$ $\mu$m, which is in good agreement with the literature value for Py waveguides~\cite{Sekiguchi2012Attenuation}.
	
	This clearly shows that the magnonic Fibonacci crystals can be considered as an effective ferromagnetic medium that can carry SWs as efficiently as regular magnonic crystals. Moreover, in the limit of long SW wavelength, the reduced order of the quasicrystal does not increase the damping of SWs or hinder their propagation.
	
	\subsection{Mini Band Gaps}
	
	Although the calculated broad band gap only spans the range from 5.0 to 5.8 GHz, we also observed no excitation of SWs in the sample at 4.6 GHz (marked C), indicating that an additional mini-band gap opens in the SW spectrum at this frequency. This agrees with our numerical calculations, where we also find additional mini-gaps. This is characteristic for waves in quasicrystals, as the spectrum is much more complex and can feature additional band gaps in comparison to the analogous periodic structure. The existence of such gaps in the spectrum (indicated by gray horizontal bars in Figure~\ref{calculation}a) is a consequence of the long-range quasiperiodic order and the self-similarity of the structure.
	
	Considering the overall band structure of the magnonic quasicrystal, it is clear that modes A and B fall into the lower band, explaining the long wavelengths of propagating SWs shown in Figure~\ref{stxm}a. Mode C lies in the mini-gap within the first band, thus, SW propagation is prohibited, corresponding to the weak SW amplitude measured in STXM at this frequency. For mode D, SW propagation is recovered as it occurs at a frequency within the band. Mode E falls in the wide main bandgap, which separates the first and the second band. At this frequency, no propagating SWs and only forced oscillation below the stripline can be observed, which is the reason for the strong decay of SW amplitude when moving away from the excitation source. Modes of even higher frequency, \textit{e.g.} mode F, lie already above the main gap and originate from the second band.
	
	This shows that the quasiperiodictiy can be used to finely tune frequencies that are forbidden, \textit{i.e.} lie in a mini band gap, and that quasicrystals offer a superior flexibility over regular magnonic crystals for band structure engineering.
	
	\subsection{Localization}
	
	\begin{figure}[t]
		\includegraphics[width=\columnwidth]{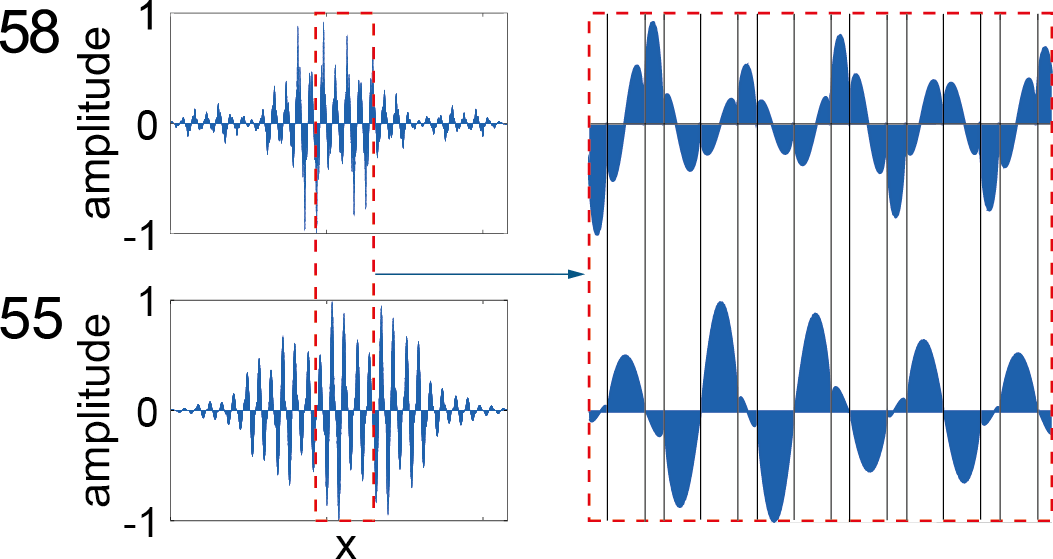}
		\caption{Comparison of amplitude distribution of SW modes in the first band (last mode, solution number 55) and second band (first mode, solution number 58). And enlargement of the central part (right panel, enclosed by red dashes) with solid grey lines marking the air gaps between the stripes. Note the nodal line within wide stripes for the second band.\label{secondband}}
	\end{figure}
	
	\begin{figure}[t]
		\includegraphics[width=\columnwidth]{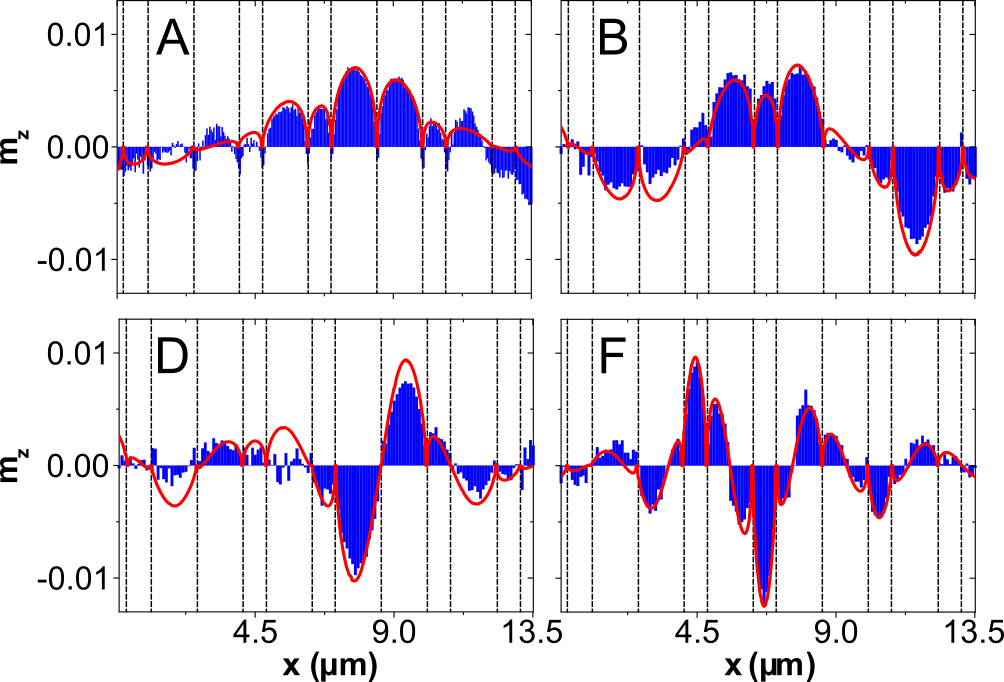}
		\caption{Calculated (red solid lines) and experimental profiles (blue bars) of allowed SW modes (\textit{cf.} Figure~\ref{stxm}a). Vertical dashed lines indicate the positions of the air gaps between the Py NWs in the structure.\label{localization}}
	\end{figure}
	
	To analyze the SW localization in the Fibonacci latice, the spatial distributions of the SW amplitude are shown in Figure~\ref{calculation}d for selected modes.
	
	The low frequency excitations are standing waves distributed throughout the whole simulated structure, whereas the number of nodes (points with a change of the sign of the amplitude) increases with frequency. The mode at the lowest frequency (solution number 1) does not have any change of the amplitude sign, the second mode has one nodal line in the middle of the structure, and so on (see the envelope of the first and third solutions presented in Figure~\ref{calculation}d). This agrees well with the expectation for regular lattices. At higher frequencies, however, the envelope of the SWs becomes irregular with several maxima located in different parts of the structure, as shown in Figure~\ref{calculation}d for solution number 20 and 43 corresponding to modes B and D respectively. Thus, they are transition modes separating the harmonic from the localized modes. This is already discussed in literature for photons~\cite{Kohmoto1987Localization} and electrons~\cite{Liu1987Electronic}, but so far not described for SWs.
	
	Nevertheless, up to a frequency of 5.0 GHz, the standing waves are solely formed from fundamental oscillations of single NWs, \textit{i.e.} without nodal points inside NWs. This property justifies grouping of these modes into one main band. Thus, the last mode (solution number 55) in the first band has antiphase oscillations in neighbouring wide NWs (\textit{cf.} Figure~\ref{secondband}). The second band, starting at 5.8 GHz (solution number 58), continues the harmonic oscillation series with irregular envelopes, but all of the excitations from this band feature a nodal line within the wide NWs, an indicative property of SWs from this band (\textit{cf.} Figure~\ref{secondband}). The same property is observed for the measured mode F originating from the second band. 
	
	In Figure~\ref{localization} a comparison between the calculated and experimental SW mode profile is shown for selected modes. The calculated results are superimposed as a red line on the experimental values shown as blue bars. Calculated profiles were adjusted to account for exponential SW decay, \textit{i.e.} the decay of the SW amplitude envelope with increasing distance from the signal line. For the lowest measured frequency in the first band (mode A), the sign of $m_z$ in all NWs is the same, except for the outmost visible stripes, which means that half of the wavelength is almost as long as the measurement window and about 10 $\mu$m. As expected, for higher frequencies shorter waves are excited. For modes B and D, the wavelength $\lambda$ is observed to be 10 $\mu$m and 4.4 $\mu$m respectively. While modes D (first band) and F (second band) both feature a wavelength $\lambda$ of 4.4 $\mu$m, mode F additionally features a SW node within a wide stripe as is expected for the second band.
	
	This indicates that the main amplitude of different SW modes can be localized in different parts of the Fibonacci quasicrystal or that the relative phase between different parts of the magnonic quasicrystals can be tuned, \textit{e.g.} between wide and narrow stripes or between both ends of wide stripes.
	
	\begin{figure*}[t]
		\includegraphics[width=0.9\textwidth]{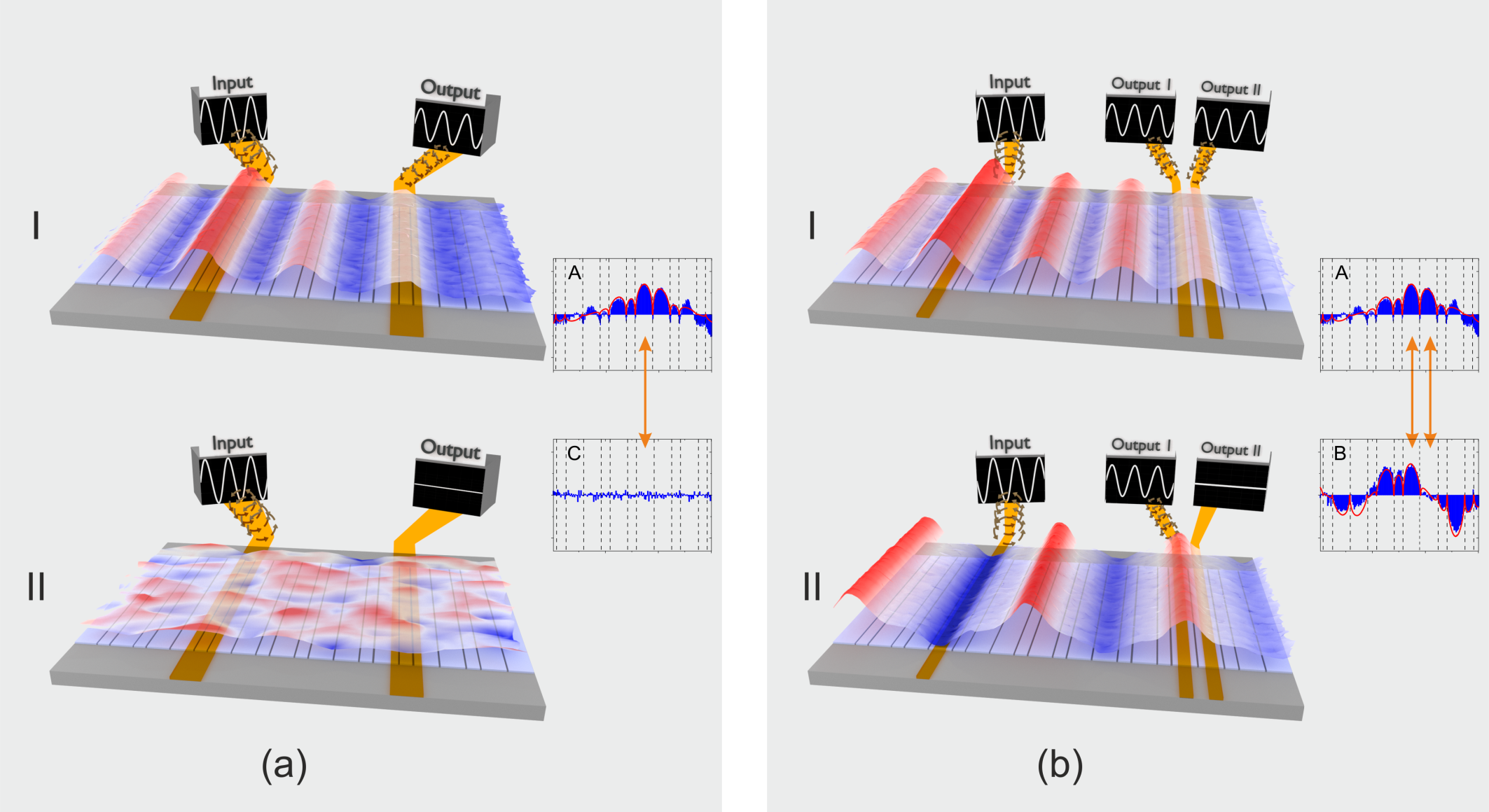}
		\caption{Artist's impression of possible magnonic devices using magnonic quasicrystals derived from the measured mode profiles shown in the insets. \emph{(a)} shows a spin wave filter with very fine granularity that uses the mini band gaps of the Fibonacci crystal to selectively damp out spin waves, where cases I and II could be realized with modes A and C. \emph{(b)} shows a device that uses the fractal localization of different spin wave modes in the quasicrystal to allow switching of the spin wave amplitude between output antennas at different positions. Such a spin wave demultiplexer with states I and II could be realized with modes A and B.\label{device}}
	\end{figure*}
	
	\subsection{Design Rules}
	
	To derive a semi-analytical description of the magnonic properties of the quasicrystal, the spatial Fourier spectrum of the spatially dependent material parameter (\textit{e.g.} $M_S$) was calculated for the considered Fibonacci $S_{11}$ sequence. The resulting 1D reciprocal lattice vectors $G$ of the Fourier components determine the SW wave vectors $k$ for which the Bragg condition ($G/2 = k$) is fulfilled. In Figure~\ref{calculation}, the blue arrows link the: (c) reciprocal lattice vectors $G$, corresponding to the Bragg peaks of highest intensity to (b) the frequencies of the SWs in the effective medium, which then point at (a) the largest magnonic gaps of the Fibonacci quasicrystal. Thus, a graphical method to determine the band structure of the quasicrystal is presented here. The integer numbers in brackets over the highest Bragg peaks at (c) are the pairs of indexes for 1D reciprocal vectors (\textit{cf.} Equation~\ref{eqa2} in the appendix) of the Fibonacci quasicrystal. At (d) the amplitude distributions of the selected modes are shown. The modes with solution number 20 and 43 are compared with the measured modes B and D respectively.
	
	This results in a simple design rule that only requires a Fourier transformation and the dispersion relation for the effective medium to design a complex and versatile band structure for magnons in quasiperiodic structures. It is easily conceivable to use the propagation, extinction in mini band gaps, and localization in tailored magnonic quasicrystals for magnonics applications. The features of the magnonic quasicrystal, \textit{i.e.} information transmission by SW propagation, SW extinction in mini band gaps, and SW localization tuning, combined with these simple design rules provide a powerful toolbox for design of magnonic devices. Figure~\ref{device} shows an artists impression of two possible devices based on our findings.
	
	In Figure~\ref{device}a a SW filter is shown that would utilize the mini band gaps in the magnonic quasicrystal to provide a finely tunable SW filter. In our samples this could be realized for modes A and C (\textit{cf.} Figure~\ref{stxm}). Spin waves at a frequency of 3.6 GHz can propagate unhindered from input to output antenna, transmitting the signal, while no SW is transmitted at 4.6 GHz, isolating the two antennas.
	
	Furthermore, the localization property of the magnonic quasicrystal could be utilized to realize a SW demultiplexer. As the SW amplitude is localized to specific areas of the Fibonacci structure, the SW could be switched between appropriately located output antennas. A rendition of such a device is shown in Figure~\ref{device}b. In our samples this could be realized for modes A and B (\textit{cf.} Figure~\ref{localization}). At 3.6 GHz two neighboring wide stripes exhibit a SW amplitude with the same phase, resulting in the same signal transmitted into hypothetical output antennas on those stripes. However, at 4.2 GHz only one of the two exhibits a SW amplitude while the other shows no intensity, thus, only one output would get activated.
	
	\section{Conclusion}
	We have experimentally observed propagating SWs in real space and time domain in 1D Fibonacci quasicrystals of dipolarly coupled Py NWs of two different sizes and fully recovered and explained this system using numerical calculations. Thereby, we have demonstrated the existence of propagating SW modes in such quasicrystals, crucial for future magnonic data processing applications. We have shown that SW propagation is not restricted to the long-wavelength limit, for which the structure can be considered as an effective medium, but also occurs at higher frequencies, for which the structure’s long-range quasiperiodic order is critical. Additionally, we have experimentally proven the existence of mini-band gaps and SW localization that are a direct consequence of the collective SW effects in magnonic quasicrystals. Furthermore, a simple analytical method has been derived for the estimation of the gaps and mini-gaps in the SW spectra of 1D quasicrystals, providing a powerful tool for designing quasiperiodic systems.
	
	The mini-gaps are wide enough to prohibit propagation of SWs despite the finite SW linewidth, thus, offering usefulness for potential applications in the filtering of microwave signals. Moreover, propagating SWs in a systems with a dense spectrum of allowed and forbidden bands offer unprecedented flexibility in the design of effective non-linear processes~\cite{Fradkin-Kashi2001Multiple}, which is one of the main challenges in the application of magnonics to transfer and process information~\cite{Chumak2014Magnon}. 
	
	\section{Acknowledgments}
	The study has received funding from the European Union’s Horizon 2020 research and innovation programme under the Marie Sk\l{}odowska-Curie GA No644348 (MagIC) and NCN Poland UMO-2012/07/E/ST3/00538.. Helmholtz Zentrum Berlin / BESSY II is gratefully acknowledged for allocating beam time at the MAXYMUS end station. J.W.K. would like to acknowledge the support of the Foundation of Alfried Krupp Kolleg Greifswald and the National Science Centre of Poland Grant No.: UMO-2016/21/B/ST3/00452. J.R. would like to acknowledge the financial support from the Adam Mickiewicz University Foundation and from the National Science Centre of Poland under ETIUDA grant number UMO-2017/24/T/ST3/00173.
	
	\appendix*
	
	\section{Estimation of the band gaps and mini-gaps positions}
	In the SW spectrum presented in Figure~\ref{calculation}a, we have distinguished two main bands separated by a wide band gap between 5.0 and 5.8 GHz, and a set of mini-band gaps of much smaller widths, both types were confirmed by STXM measurements. The formation of these gaps can be related to the Bragg scattering of SWs with wavenumbers fulfilling the Bragg condition. Thus, the location of the 1D reciprocal lattice vectors $G$ in the reciprocal space should give us information about the position of the frequency gaps and mini-gaps in the spectrum. The procedure described below allows connecting the structure of the magnonic quasicrystal with the frequency gaps in the SW spectrum. 
	
	For a wave with wavenumber $k$, the Bragg condition may be written as:
	\begin{equation}
	\frac{G(h_1,h_2)}{2}=k,\label{eqa1}
	\end{equation}
	where $h_1$ and $h_2$are integer numbers. The 1D reciprocal lattice vectors G($h_1,h_2$) are defined for the Fibonacci lattice as~\cite{Limonov2012Optical,Janot2012Quasicrystals}:
	\begin{equation}
	G(h_1,h_2 )=\frac{2\pi}{\bar{a}}(h_1+\frac{1}{\varphi}h_2),\label{eqa2}
	\end{equation}
	where:
	\begin{equation}
	\bar{a} =\frac{\varphi(W_W+W_G)+(W_N+W_G)}{\varphi +1}\label{eqa3}
	\end{equation}
	is an averaged period of the Fibonacci structure, and $\varphi=\frac{1+\sqrt{5}}{2}$ is the so called golden ratio. The 1D reciprocal lattice numbers (\ref{eqa2}) are indexed by two integer numbers $h_1$ and $h_2$. To explain this we recall that the Fibonacci sequence can be formed by the projection of a square lattice to a straight line tilted at an irrational angle $\alpha$, where $\tan(\alpha)=1/\varphi$. The projection is done from the stripe parallel to the mentioned line~\cite{Janot2012Quasicrystals}. To obtain the reciprocal lattice numbers (\ref{eqa2}) for the corresponding quasiperiodic sequence of lattice nodes, obtained from the projection to the line inclined at an angle $\alpha$,  the Fourier transform of a square lattice (set of Dirac delta peaks) and the Fourier transform of the stripes (sinc function) have to be convoluted~\cite{Janot2012Quasicrystals}. Note that for 1D Fibonacci structures, the 1D reciprocal lattice vectors are densely packed numbers in 1D reciprocal space, due to the irrational factor $1/\varphi$ in Equation~\ref{eqa2}. The wide gap is related to the Bragg scattering of SWs at the smaller 1D reciprocal vectors $G(h_1=0,h_2=0)$. For the Py NW Fibonacci structure considered here, $\bar{a}=1212.6$ nm and the main Bragg resonance condition (\ref{eqa1}) is fulfilled for $k = \SI{1.6}{\per\micro\meter}$ which is close to the wavenumber of the experimentally detected SWs for one of the first modes from the second band at the edge of the main gap ($\SI{1.42}{\per\micro\meter}$ for the \SI{5.7}{\giga\hertz}, mode F). The gaps of smaller width are indicated by 1D reciprocal lattice vectors $G$ determined by other values of indices $h_1$ and $h_2$.
	
	To predict the position of the gap frequencies in 1D quasicrystals we need to project the wavenumbers found from Equation~\ref{eqa1} onto the dispersion relation of SWs. Therefore we propose to use the dispersion relation of a homogeneous ferromagnetic film with effective parameters. The measured and calculated SW profiles (Figure~\ref{localization}) show a strong reduction of the dynamic components of the magnetization amplitude close to the edges of the NWs. This proves a significant dynamical demagnetizing effect, induced by the presence of NW edges, which leads to pinning of the dynamical magnetization and to an increase of the FMR frequency. Thus, the FMR frequency for the patterned magnetic system considered here with an in-plain shape anisotropy can be expressed with the demagnetizing factors $N_d$ and $1-N_d$ which take values from 0 to 1)~\cite{Stancil2009Spin}:
	\begin{equation}
	f_\text{FMR}=\frac{1}{2\pi}\sqrt{(\omega_0+N_d\omega_M)(\omega_0+(1-N_d)\omega_M)}\label{eqa4}
	\end{equation}
	where $\omega_0=\gamma\mu_0 H_\text{app}$ and $\omega_M=\gamma\mu_0 M_S$ are characteristic frequencies corresponding to the applied field and the saturation magnetization (expressed in frequency units) respectively.
	
	$\mu_0$ and $\gamma$ are the magnetic permeability of vacuum and the gyromagnetic ratio respectively. For the same parameters as in the numerical simulations, we were able to shift the FMR frequency $f_\text{FMR}$ (\ref{eqa4}) to the value obtained in numerical calculations (3.6 GHz) by fitting the demagnetizing factor $N_d$. The relatively small value of $N_d=0.0157$ ensures the required upshift of $f_\text{FMR}$.
	
	The dispersion relation for the homogeneous film with effective material parameters and in-plane demagnetizing effects can be calculated according to:
	\begin{widetext}
		\begin{equation}
		f(k)=\frac{1}{2\pi} \sqrt{(\omega_0+\omega_M\lambda^2k^2+N_d\omega_M)(\omega_0+\omega_M\lambda^2k^2+(N_d-1)\omega_M)+\omega_M^2/4(1-e^{-2kd})}\label{eqa5}
		\end{equation}
	\end{widetext}
	with the exchange length $\lambda=\sqrt{\frac{2A}{\mu_0 M_S^2 }}=6.19$ nm for Py. Using the dispersion relation (\ref{eqa5}), we estimate the maximum value of the wavenumber $k_\text{max}=\SI{0.0035}{\per\nano\meter}$ for the frequency range up to 6.5 GHz considered in numerical calculations. The obtained dispersion relation is shown in Figure~\ref{calculation}b. By projecting the wavenumbers fulfilling the Bragg condition (\ref{eqa1}) on the dispersion relation $f(k)$ we are able to determine the expected positions of the band gaps, as shown in Figure~\ref{calculation}b.
	
	Apart from the frequencies for which gaps open, we can also estimate the relative widths of these gaps. This can be done with the help of the diffraction spectra of the structure. The diffraction properties of any kind of structure are given by the structure factor $F(G)$ which can be estimated by Fourier transform of the spatial dependence of the material parameter, \textit{e.g.} the saturation magnetization $M_S$, in the composite system. We calculated the discrete Fourier transform of $M_S$:
	\begin{equation}
	F(G_l)\propto \sum_{j=0}^{N-1}M_S(x_j)e^{-iG_lx_j},\label{eqa6}
	\end{equation}
	where $x_j=L j/N$ is the location of $M_S (x_i )$ in the sample in real space and $G_l=\frac{2\pi}{L}l$ is the discretized reciprocal lattice number. The modulus $|F(G_l)|$ obtained for the structure considered here is presented in Figure~\ref{calculation}c. The height of the bars representing $|F(G_l)|$ is related to the intensity of the Bragg peaks of the scattered waves. The location $G_l$  of Bragg peaks with large intensity corresponds to the frequency gaps opened at $f_\text{gap}$ given by $k(f_\text{gap})=G_l/2$, for which the relation $k(f)$ can be found from the dispersion relation (\ref{eqa5}). The position of the highest peaks from the numerically calculated Fourier spectrum of the quasiperiodic structure strictly coincides with the wavenumbers fulfilling the analytical formula (\ref{eqa2}). 
	
	
	\bibliography{article}
	
\end{document}